\documentclass[pra,showpacs,superscriptaddress,preprint]{revtex4} 

\usepackage{amsfonts}
\usepackage{amsmath}
\usepackage{amssymb}
\usepackage{graphicx}
\usepackage{epsfig}
\usepackage{bm}

\begin{document}

\preprint{APS}
\title{Unambiguous modification of nonorthogonal single and two photon polarization states}
\author{F. A. Torres-Ruiz}
\affiliation{Center for Quantum Optics and Quantum Information,
Departamento de F\'{\i}sica, Universidad de Concepci\'{o}n,
Casilla 160-C, Concepci\'{o}n, Chile.}
\author{J. Aguirre}
\affiliation{Center for Quantum Optics and Quantum Information,
Departamento de F\'{\i}sica, Universidad de Concepci\'{o}n,
Casilla 160-C, Concepci\'{o}n, Chile.}
\author{A. Delgado}
\affiliation{Center for Quantum Optics and Quantum Information,
Departamento de F\'{\i}sica, Universidad de Concepci\'{o}n,
Casilla 160-C, Concepci\'{o}n, Chile.}
\author{G. Lima}
\affiliation{Center for Quantum Optics and Quantum Information,
Departamento de F\'{\i}sica, Universidad de Concepci\'{o}n,
Casilla 160-C, Concepci\'{o}n, Chile.}
\author{L. Neves}
\affiliation{Center for Quantum Optics and Quantum Information,
Departamento de F\'{\i}sica, Universidad de Concepci\'{o}n,
Casilla 160-C, Concepci\'{o}n, Chile.}
\author{S. P\'{a}dua}
\affiliation{Center for Quantum Optics and Quantum Information,
Departamento de F\'{\i}sica, Universidad de Concepci\'{o}n,
Casilla 160-C, Concepci\'{o}n, Chile.} \affiliation{Departamento
de F\'{\i}sica, Universidade Federal de Minas Gerais, Caixa Postal
702, 30123-970 Belo~Horizonte~MG, Brazil.}
\author{L. Roa}
\affiliation{Center for Quantum Optics and Quantum Information,
Departamento de F\'{\i}sica, Universidad de Concepci\'{o}n,
Casilla 160-C, Concepci\'{o}n, Chile.}
\author{C. Saavedra}
\affiliation{Center for Quantum Optics and Quantum Information,
Departamento de F\'{\i}sica, Universidad de Concepci\'{o}n,
Casilla 160-C, Concepci\'{o}n, Chile.}
\email{carlos.saavedra@udec.cl}
\date{\today}

\begin{abstract}
In this work we propose a probabilistic method which allows an unambiguous
modification of two non-orthogonal quantum states. We experimentally implement
this protocol by using two-photon polarization states generated in the process
of spontaneous parametric down conversion. In the experiment, for codifying initial
quantum states, we consider single photon states and heralded detection. We show that
the application of this protocol to entangled states, it allows a fine control of the
amount of entanglement of the initial state.
\end{abstract}

\maketitle


\newpage

\section{Introduction}
\label{sec:intro}

The discrimination of non-orthogonal quantum states has been studied extensively.
Main motivations driving studies in this problem are the fact that quantum states
are not observables \cite{Peres0} and the possibility of encoding information on
states of quantum systems \cite{Nielsen}. Seminal work on this problem started with
the studies of Helstrom and Holevo on quantum state identification and probability
theory \cite{Helstrom_Holevo} which led to the minimum error discrimination strategy.
In this strategy the possible input states, pure or mixed, are identified with some error.

A different strategy, unambiguous state discrimination (USD), has been studied for
several families of pure states \cite{Ivanovic,Dieks,Peres,Jaeger,Chefles,Chefles2}.
Here, the states to be discriminated are perfectly  identified but with the addition
of an inconclusive event. The first experiment for unambiguous state discrimination
of two non-orthogonal polarization states of light near the Ivanovic-Dieks-Peres (IDP)
limit was implemented via weak optical pulses propagated through an optical fiber
with polarization-dependent loss \cite{Huttner}. USD at the IDP limit was experimentally
achieved via a free-space interferometer which recorded conclusive and inconclusive events
\cite{Clarke} and has also been experimentally studied in the context of
three non-orthogonal linearly independent states \cite{Mohseni}.
Recently, a proposal for doing the conclusive discrimination of $2^{M}$ (with $M$ integer)
symmetric states, considering only linear optics, was reported \cite{Jimenez}. Today, USD
plays an important role in quantum communication and quantum computing,
being at the core of many quantum cryptographic schemes and probabilistic quantum algorithms
\cite{BergouRev}. It can be used, for instance, as a model for efficient attack
in quantum cryptography \cite{Fuchs,Dusek}.

USD can be understood as a probabilistic, conclusive mapping of a
family of linearly independent non-orthogonal states onto a set of orthogonal ones.
Thus, it is also possible to conceive probabilistic, conclusive mappings between sets of
linearly independent, non-orthogonal states. In this article, we study and experimentally
implement the case of probabilistically and conclusively mapping two non-orthogonal initial states
${|\alpha_{\pm}\rangle}$ onto two non-orthogonal final states ${|\beta_{\pm}\rangle}$.
Our goal is to map the initial states onto the final states with a prescribed inner product.
In this context we refer to the mapping as conclusive modification of the inner product (CMIP).
Initial and final states are assumed to be known, that is, the base in which they are generated
and the coefficients of the expansion of the states in this base are known. Furthermore,
we allow for generating the two initial states with arbitrary a priori probabilities.
This condition forbids the use of unitary transformations to implement the mapping since
those preserve the inner product. A particular class of the probabilistic mapping that
we study here has been previously introduced in connection with the probabilistic,
state dependent cloning machine \cite{Duan}.

In the experimental implementation of the CMIP we use the polarization degree of freedom of photons.
These are generated in pairs in the spontaneous parametric down conversion (SPDC) process and are
selected in a factorized polarization state. One photon of the pair is used as a trigger,
whilst the other is used for codifying two initial polarization states whose inner product
is conclusively modified. The mapping is implemented with conditional operations applied
onto the polarization. These operations are dependent on the propagation path of the photon.
As an application of the CMIP, we show that, when the transformation is applied to a partially
entangled two-photon polarization state, it allows a fine control of its amount of entanglement.
In particular, this allows one to filter maximally entangled states.

This article is organized as follows: In Sec.~\ref{sec:theory} we describe the sequence of
conditional operations that physically implements the CMIP. In Sec.~\ref{sec:experiment}
we report experimental results based on single-photon states which are obtained by generating
factorized two-photon states and by considering heralded detection. In Sec.~\ref{sec:apply} we apply the CMIP onto one of the down-converted photons in a partially entangled state. This case is also experimentally implemented. Finally, we summarize and conclude in Sec.~\ref{sec:summary}.

\section{Conclusive modification of the inner product}
\label{sec:theory}

Let us consider that the initial single photon state is described by one of two possible superpositions, $|\psi^{(+)}\rangle $ or $|\psi ^{(-)}\rangle $, defined by \begin{equation}\label{eq:psipm}
|\psi ^{(\pm) }\rangle =\cos (\alpha/2)| H_1 \rangle \pm \sin( \alpha /2)| V_1 \rangle ,
\end{equation}
where $| H_1 \rangle $ and $| V_1 \rangle$ denote horizontal and vertical polarization, respectively.  $\alpha$ is the angle between the linear polarizations and the subindex denotes the propagation path. As an ancillary system we consider two effective distinguishable propagation paths, $| 1 \rangle$ and $| 2 \rangle$  \cite{Cerf98}, so that the state of the photon with polarization $k$ ($k=H,V$) propagating along path $j$ ($j=1,2$) is denoted by $|k_j\rangle$.

Here, the photon is assumed to be initially propagating along the path $| 1 \rangle$.
After the application of a conditional operation, the initial states $|\psi^{(+)}\rangle $ and
$|\psi^{(-)}\rangle $ will probabilistically have the angle between them modified. The angle $\alpha$ of the initial states will be changed to $\beta$. The final states $|\phi ^{(\pm)}\rangle $ are given by
\begin{equation}  \label{eq:phi}
|\phi ^{(\pm)}\rangle =\cos(\beta/2)| H_1 \rangle \pm \sin(\beta/2) | V_1 \rangle.
\end{equation}

\begin{figure}[t]
\centering{\rotatebox{-90}{\includegraphics[width=0.325\textwidth]{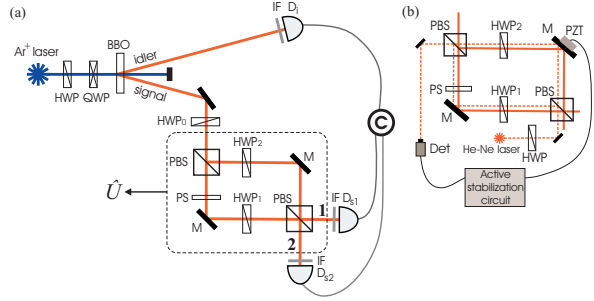}}}
\vspace{-1.5cm}
\caption{(a) Sketch of the experimental setup. Photon pairs are generated by SPDC in a type-II (doube type-I) BBO crystal for the experiment described in Sec.~\ref{sec:experiment} (Sec.~\ref{sec:apply}). The interferometer in the dashed box performs the CMIP protocol. HWP, half-wave plate; QWP, quarter-wave plate; M, mirror; PBS, polarizing beam splitter; PS, phase shifter; IF, interference filter; D$_j$ ($j = i, s1, s2$), single photon detectors; C, single and coincidence counter. (b) Active stabilization circuit of the interferometer for the experiment in Sec.~\ref{sec:apply}. PZT, piezoelectric actuator; Det, detector.}
\label{fig:setup}
\end{figure}

The experimental setup for implementing this conditional operation is depicted in the dashed box of Fig.~\ref{fig:setup}(a). This is based on a balanced Mach-Zehnder like interferometer, where polarizing beam splitters (PBS) have been used both for generating different propagation paths and for recombining them. Half-wave retardation plates (HWP) inserted in both interferometer paths modify the horizontal (vertical) polarization when the condition $0\leq \alpha < \beta \leq \pi $ ($0\leq \beta < \alpha \leq \pi $) holds. After the first PBS, each polarization propagates along a different path, so that conditional operations can be applied. The phase difference arising due to the interferometer arms' length difference, in the coherence region, is modified by rotating a thin piece of glass inserted in the interferometer arms (PS). The second PBS combines the polarization components.

In the first case, we increase the angle between the initial states, i.e.,
$0\leq \alpha \leq \beta \leq \pi $. In particular, the case of USD (initial states going to orthogonal states) occurs when $\beta=\pi/2$. We perform a conditional unitary operation, $\hat{U}$, which acts in a subspace of the composite system spanned by $\{| H_1 \rangle,| V_1 \rangle,| V_2 \rangle\}$. Experimentally this is done by rotating the HWP$_1$ at the propagation path $|1\rangle$ while HWP$_2$ is kept fixed at zero angle. The action of this conditional operation is such that
\begin{eqnarray}
\label{eq:U1}
\hat{U} | H_1 \rangle & = & \cos (2\gamma_1) | H_1 \rangle +\sin (2\gamma_1) | V_2 \rangle,
\nonumber\\
\hat{U} | V_1 \rangle & = & e^{i\varphi}   | V_1 \rangle.
\end{eqnarray}
$ \varphi$ accounts for a phase difference coming through the propagation in different paths, and $\gamma_1$ is the rotation angle of HWP$_1$. If we consider that $\gamma_1$ and the angles $\alpha$ and $\beta$ satisfy the relation
\begin{equation}
\gamma_1 =\frac{1}{2} \arccos \left(\frac{\tan (\alpha /2)}{\tan (\beta /2)}\right),
\label{gamma1}
\end{equation}
and that $\varphi=0$, the initial state of the composite system in Eq.~(\ref{eq:psipm}) will evolve according to
\begin{equation}
\hat{U}|\psi^{(\pm)}\rangle = \sqrt{P_{\mathrm{suc}}}
|\phi^{(\pm)}\rangle +\sqrt{1-P_{\mathrm{suc}}}| V_2 \rangle,
\label{U1}
\end{equation}
where $P_{\mathrm{suc}}$ is the success probability for the conclusive modification of the
inner product, which is obtained when the photon is propagating along path $|1\rangle$.
This probability is given by
\begin{equation}
\label{eq:psgt}
P_{\mathrm{suc}}(\alpha,\beta)=\frac{\sin ^{2}(\alpha /2)}{\sin ^{2}(\beta /2)},
\end{equation}
which applies when $0\leq \alpha \leq \beta \leq \pi $. The failure case occurs when,
after the conditional operation, the photon has vertical polarization
and propagates along path $|2\rangle$. In other words, the conditional unitary operation
followed by a projection measurement onto the path ancillary system allows one to transform, with probability $P_{\mathrm{suc}}$, the two states $|\psi^{(\pm)}\rangle $ with angle $\alpha$ onto the $|\phi^{(\pm)}\rangle$ states, respectively, which have an inner angle $\beta \ge \alpha$. We would like to remark that this scheme can be used to prepare two orthogonal states starting from two non-orthogonal ones. This occurs when $\beta=\pi/2$ and it is achieved with probability $P_{\mathrm{suc}}(\alpha,\pi/2)=2\sin ^{2}(\alpha /2)$, which is the optimal probability of unambiguous discrimination of two non-orthogonal quantum states
\cite{Peres}.

In the second case, we consider $0\leq \beta \leq \alpha $, which corresponds to the case where the initial $|\psi^{(\pm)}\rangle $ states go to the final $|\phi^{(\pm)}\rangle $ states with a smaller inner angle. This can be directly done by exchanging the roles of the HWP$_1$ and HWP$_2$ in the interferometer. The conditional operation being performed can now be written as
\begin{eqnarray}
\label{eq:U2}
\hat{U}'| H_1 \rangle &=& e^{i\varphi'}   | H_1 \rangle \nonumber \\
\hat{U}'| V_1 \rangle &=& \cos (2\gamma_2) | V_1 \rangle + \sin (2\gamma_2) | H_2 \rangle,
\end{eqnarray}
where $\gamma_2$ is the rotation angle of the HWP$_2$.
Therefore, the two incoming states $| \psi^{(\pm)} \rangle $ are transformed as follows:
\begin{equation}
\hat{U}' |\psi^{(\pm)} \rangle =
\sqrt{P_{\mathrm{suc}}'} | \phi^{(\pm)}\rangle \pm
\sqrt{1-P_{\mathrm{suc}}'}| H_2 \rangle, \label{U2}
\end{equation}
where we choose the parameters of the transformation to be
\begin{equation}
\gamma_2 = -\frac{1}{2} \arccos \left(- \frac{\tan (\beta /2)}{\tan (\alpha /2)}\right),
\label{gamma2}
\end{equation}
and $\varphi'=0$. The minus sign appears now because we have considered that the fast axis of the HWP$_2$ is at the horizontal direction when $\gamma_2 = 0$. In this case the success probability is given by
\begin{equation}
P_{\mathrm{suc}}'(\alpha,\beta)=\frac{\cos ^{2}(\alpha /2)}{\cos ^{2}(\beta /2)},
\label{eq:psst}
\end{equation}
which is valid for the case of $0\leq \beta \leq \alpha $. The failure case occurs when, after the device, the photon with horizontal polarization is propagating along path $|2\rangle$, with probability $1-P_{\mathrm{suc}}'$. We notice that the probability of generating non-orthogonal states starting from two orthogonal ones is $P_{\mathrm{suc}}'(\pi/2,\beta)=1/[2\cos ^{2}(\beta /2)]$. This probability is always higher than the probability of generating two orthogonal states which is $P_{\mathrm{suc}}(\alpha,\pi/2)=2\sin ^{2}(\alpha /2)$.

Here, we also want to remark that although we have considered a single system (one photon) for defining both the physical and the ancillary systems, the above description also applies for two-party quantum states as it will be described in Sec. \ref{sec:entanglement}. For instance, this can be applied in the context of cavity quantum electrodynamics where the system is described by one mode of the electromagnetic field, resonant to a high Q cavity, and the ancilla by a two-level atom \cite{Davidovich94,Brune99}.

\section{Experimental results}
\label{sec:experiment}

The experimental setup for demonstrating the CMIP is sketched in Fig.~\ref{fig:setup}(a). A
351.1 nm single-mode Ar-ion laser pumps with 200 mW a $5$-mm-thick $\mathrm{BBO}$ ($\beta $-Barium Borate) crystal, cut for type-II phase matching. Pairs of photons, usually called signal (s) and idler (i), are generated non-collinearly by SPDC and those with the same wavelength of 702.2 nm are selected by 10.0 nm bandwidth Gaussian interference filters centered at this wavelength and placed in front of the photo-detectors. We select the photon pairs in a separable polarization state given by $|H\rangle_{s}|V\rangle _{i}$. The idler goes directly to $D_i$ detector and its detection heralds the presence of a photon in the signal arm to perform the CMIP protocol. The half-wave plate (HWP$_0$) before the interferometer prepares one of the input states given by Eq.~(\ref{eq:psipm}). After the interferometer, the signal photon is detected at output port 1 by $D_{s1}$. Single and coincidence counts between $D_i$ and $D_{s1}$ are registered in a counter (C) with a resolving time of 5~ns. The coincidence events ensure that the protocol has successfully been performed onto the signal photon. The quality of the experimental setup was tested by measuring the visibility of the single-photon interference pattern at the interferometer. For this purpose the polarization of the input signal photon was rotated to $\pi/4$ and a linear polarizer rotated at an angle of $\pi/4$ was inserted in front of the detector $D_{s1}$. Interference pattern with visibility of $90 \%$ was recorded by modifying the length of one of the interferometer's arm.

\begin{figure}[t]
\centering{\rotatebox{-90}{\includegraphics[width=0.35\textwidth]{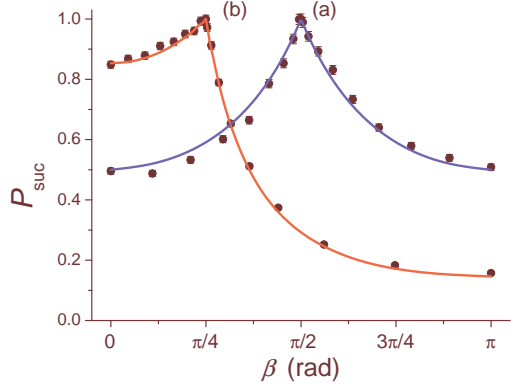}}}
\caption{Experimental success probabilities (black points) as a function of the output inner angle $\beta $. The initial superposition was $|\psi _{\pm }\rangle $, with (a) $\alpha =\pi /2$ and (b) $\alpha =\pi/4$. The solid curve corresponds to the success probability $P_{\mathrm{suc}}$ [Eq.~(\ref{eq:psgt})] for $\beta $ satisfying $\alpha \leq \beta \leq \pi$ and $P_{\mathrm{suc}}'$ [Eq.~(\ref{eq:psst})] for $\beta $ satisfying $0\leq \beta \leq \alpha $.}
\label{fig:probabilities}
\end{figure}

In the first test of the protocol, we have unambiguously prepared non-orthogonal quantum states from initially orthogonal ones, i.e., with $\alpha=\pi/2$ in Eq.~(\ref{eq:psipm}).
For the interval $0\leq\beta\leq\pi/2$, the coincidence rate between $D_{i}$ and $D_{s1}$ was measured as a function of the HWP$_2$ angle, $\gamma_{2}$, which is varied from $0$ up to $\pi/4$ while $\gamma _{1}=0$. For the interval $\pi/2<\beta\leq\pi$, the coincidence rate between $D_{i}$ and $D_{s1}$ was measured as a function of the HWP$_1$ angle, $\gamma_{1}$, which is varied from $0$ up to $\pi/4$ while $\gamma _{2}=0$. The success probability of generating states $|\phi^{(\pm)}\rangle $ is plotted in Fig.~\ref{fig:probabilities}(a). In the first (second) interval, the measured success probability $P_\mathrm{suc}'$ ($P_\mathrm{suc}$) is the ratio between the coincidence counts as a function of $\gamma _{2}$ ($\gamma _{1}$) and the coincidence counts for $\gamma _{1}=\gamma _{2}=0$, which is the configuration where the interferometer does not act onto the single photon state. The theoretical expression using Eq.~(\ref{eq:psst}) [Eq.~(\ref{eq:psgt})] is in good agreement with the experimental results.

\begin{figure}[t]
\centering{\includegraphics[width=0.35\textwidth]{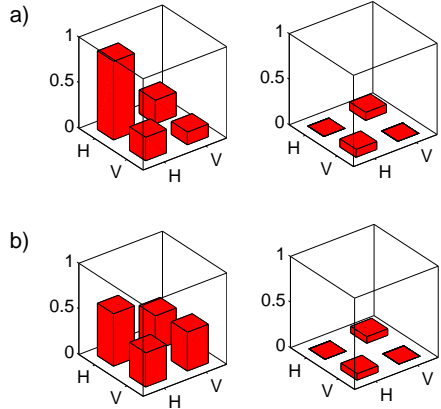}}
\caption{Tomographic reconstruction of the input state $|\psi_{+}\rangle$ with $\alpha=\pi/4$ (a) and the output state $|\phi_{+}\rangle$ with $\beta=0.44\pi$ (b). The fidelity with the expected states are $F=0.94 \pm 0.03$ and $F=0.88 \pm 0.04$, respectively. The plot at the left (right) hand side corresponds to the real (imaginary) part of the reconstructed density matrix.}
\label{fig:tomography}
\end{figure}

As a second test, we start with non-orthogonal input states with $\alpha=\pi/4$ in Eq.~(\ref{eq:psipm}), and repeat the procedure described above in order to measure the success probability of generating output states $|\phi^{(\pm)}\rangle $. For the interval $0 < \beta < \pi/4$ ($\pi/4 < \beta <\pi$ we measure $P'_{suc}$ ($P_{suc}$) by varying $\gamma_2$ and keeping $\gamma_1 = 0$ ($\gamma_1$ and keeping $\gamma_2 = 0$). In this case, experimental success probabilities of generating states $|\phi^{(\pm)}\rangle $ appears as black points in Fig.~\ref{fig:probabilities}(b). Again, solid line is obtained by using Eq.~(\ref{eq:psst}) [Eq.~(\ref{eq:psgt})], which is in excellent with the experimental results. To demonstrate the experimental modification of the inner product, we have performed the tomographic reconstruction of the polarization states \cite{James} of the input and one of the output states in case of successful measurement. The result is shown in Fig.~\ref{fig:tomography}. For the input state shown in Fig.~\ref{fig:tomography}(a) the fidelity with the state $|\psi^{(+)}\rangle$ in Eq.~(\ref{eq:psipm}) with $\alpha = \pi/4$ is $F=0.94\pm 0.03$. For the output state shown in Fig.~\ref{fig:tomography}(b) the fidelity with the state $|\phi^{(+)}\rangle$ in Eq.~(\ref{eq:phi}) with $\beta=0.44\pi$ is $F=0.88\pm 0.04$.

\section{Applications}
\label{sec:apply}

\subsection{Conclusive modification of entanglement}
\label{sec:entanglement}

In the experiment described above the photon pair has been generated in a separable state,
and one of them served as witness to the other. If this two-photon state is replaced by a
partially entangled one, then it is possible to probabilistically increase or decrease its amount of entanglement \cite{Caves03} by feeding the interferometer of Fig.~\ref{fig:setup} with one photon of the entangled pair, as we will show now. Suppose that the following two-photon state is prepared
\begin{equation}
|\Psi\rangle=\cos (\alpha/2) |H\rangle_s |H\rangle_i+ e^{i\delta} \sin (\alpha/2) |V\rangle_s |V\rangle_i.
\label{eq:partial1}
\end{equation}
Its degree of entanglement is $E(\Psi)= |\sin(\alpha)|$ \cite{Wootters}. This state may also be rewritten in the following form:
\begin{equation}
|\Psi\rangle=\frac{1}{\sqrt{2}}
\left(
|\psi^{(+)}\rangle_s |+\rangle_i+|\psi^{(-)}\rangle_s |-\rangle_i
\right),
\label{eq:partial}
\end{equation}
where $| \psi^{(\pm)} \rangle_s = \cos (\alpha/2) |H\rangle_s\pm e^{i\delta} \sin (\alpha/2) |V\rangle_s $ and $| \pm \rangle_i = (|H\rangle_i \pm |V\rangle_i)/\sqrt{2}$. The inner product $\langle\psi^{(+)} |\psi^{(-)}\rangle=\cos\alpha$. The degree of entanglement of $|\Psi\rangle$ is $E(\Psi)=|\sin(\alpha)|$. This relationship shows that by modifying the inner product of $|\psi^{(\pm)}\rangle_s$, we manipulate the entanglement of  $|\Psi\rangle$, and this can be done through the protocol developed in Sec.~\ref{sec:theory}.

For performing the CMIP, we assume the most general action of the interferometer, i.e., HWP$_1$ and HWP$_2$ in Fig.~\ref{fig:setup} are rotated at $\gamma_1$ and $\gamma_2$ angles, respectively. If it is fed with the signal photon, the two-photon state in Eq.~(\ref{eq:partial}) is transformed into
\begin{equation}
|\Phi\rangle=\sqrt{N_1}|\Phi^{(1)}\rangle - i \sqrt{N_2}|\Phi^{(2)}\rangle,
\end{equation}
where the states $|\Phi^{(1)}\rangle$ and $|\Phi^{(2)}\rangle$ are two-photon polarization states associated with the propagation paths $1$ and $2$ of the signal photon at the outputs of the interferometer, with probabilities $N_1$ and $N_2$, respectively, which are given by:
\begin{eqnarray} \label{N1}
N_1&=& \cos^2(\alpha/2) \cos^2(2\gamma_1)+\sin^2(\alpha/2)\cos^2(2\gamma_2), \label{Eq:N1}\\
N_2&=&1-N_1. \label{Eq:N2}
\end{eqnarray}
The states $|\Phi^{(1)}\rangle$ and $|\Phi^{(2)}\rangle$ are given by
\begin{eqnarray}
|\Phi^{(1)}\rangle&=&\frac{1}{\sqrt{2}}(|\phi^{(+)}_1\rangle_s|+\rangle_i+|\phi^{(-)}_1\rangle_s|-\rangle_i),
\nonumber\\
|\Phi^{(2)}\rangle&=&\frac{1}{\sqrt{2}}(|\phi^{(+)}_2\rangle_s|+\rangle_i+|\phi^{(-)}_2\rangle_s|-\rangle_i).
\label{Finaltwophotonstates}
\end{eqnarray}
Each one of these states can be filtered by distinguishing the propagation path of the signal photon at the output of the interferometer. The $\phi^{(\pm)}_1$ and $\phi^{(\pm)}_2$ states are given by:
\begin{widetext}
\begin{eqnarray}
|\phi^{(\pm)}_1\rangle_s  &=& \frac{1}{\sqrt{N_1}}[\cos (\alpha/2)\cos(2\gamma_1)|H_1\rangle_s \pm e^{i\delta}\sin (\alpha/2)\cos(2\gamma_2) |V_1\rangle_s ]   ,
\nonumber\\
|\phi^{(\pm)}_2\rangle_s &=& \frac{1}{\sqrt{N_2}}[\cos (\alpha/2)\sin(2\gamma_1) |V_2\rangle_s \pm  e^{i\delta}\sin (\alpha/2)\sin(2\gamma_2)|H_2 \rangle_s ].
\label{Finalonephotonstates}
\end{eqnarray}
\end{widetext}
The entanglement of $|\Phi^{(1)}\rangle$ and $|\Phi^{(2)}\rangle$ depends on the entanglement $E(\Psi)$ of the initial state $|\Psi\rangle$ and on the values of the rotation angles $\gamma_1$ and $\gamma_2$, and it can be written as:
\begin{eqnarray}
E(\Phi^{(1)})&=&E(\Psi) \frac{|\cos(2\gamma_1)\cos(2\gamma_2)|}{N_1} \label{eq:E1},\\
E(\Phi^{(2)})&=&E(\Psi) \frac{|\sin(2\gamma_1)\sin(2\gamma_2)|}{N_2}\label{eq:E2}.
\end{eqnarray}
We want to remark that probabilities $N_1$ and $N_2$ are functions of $E(\Psi)$, due to their dependence on $\alpha$ angle. Let us analyze these equations in the range $[0,\pi/4]$ that the HWP angles $\gamma_1$ and $\gamma_2$ can be rotated.  In case of an initial separable state, that is $E(\Psi)=0$, Eqs. (\ref{eq:E1}) and (\ref{eq:E2}) indicate that the two-photon states $|\Phi^{(1)}\rangle$ and $|\Phi^{(2)}\rangle$ are also separable. This holds for any values of the angles $\gamma_1$ and $\gamma_2$ and is consistent with the fact that local operations cannot create entanglement.

For an arbitrary initial entangled state $|\Psi\rangle$, the state $|\Phi^{(1)}\rangle$ has the same amount of entanglement of the initial state when $\gamma_1=\gamma_2=0$. Under this condition the generation probability $N_1$ becomes unitary indicating that photons entering the interferometer propagate on path 1 only. Furthermore, Eqs. (\ref{Finaltwophotonstates}) and (\ref{Finalonephotonstates}) show that states $|\Psi\rangle$ and $|\Phi^{(1)}\rangle$ become equal. A similar behavior arises when $\gamma_1=\gamma_2=\pi/4$, in this case states $|\Psi\rangle$ and $|\Phi^{(2)}\rangle$ become equal. When $\gamma_1=\gamma_2\neq 0$ or $\pi/4$ we obtain that the states $|\Phi^{(1)}\rangle$ and $|\Phi^{(2)}\rangle$ are generated with distinct probabilities but have the same entanglement degree of the initial state, that is, $E(\Phi^{(1)})=E(\Phi^{(2)})=E(\Psi)$.

Let us now consider an initial partially entangled state and search for conditions on $\gamma_1$ and $\gamma_2$ under which the entanglement $E(\Phi^{(1)})$ is greater than the entanglement $E(\Psi)$. From Eqs. (\ref{Eq:N1}) and (\ref{eq:E1}) and the condition $E(\Phi^{(1)})\ge E(\Psi)$ we obtain that the rotation angles $\gamma_1$ and $\gamma_2$ must satisfy the condition
\begin{equation}
[cos(2\gamma_1)-cos(2\gamma_2)]^2-cos(\alpha)[cos(2\gamma_2)^2-cos(2\gamma_1)^2
]\leq 0, \label{eq:increasing}
\end{equation}
for $\gamma_1$ and $\gamma_2$ in the $[0,\pi/4]$ interval. This is, for a fixed $\gamma_2$ the entanglement of $|\Phi^{(1)}\rangle$ state is increased when rotating HWP$_1$ from $\gamma_1=\gamma_2$ up to a value given by Eq.~(\ref{eq:increasing}).

Finally we study under which condition the initial arbitrary partially entangled state $|\Psi\rangle$ can be transformed into a maximally entangled state, that is $E(\Phi^{(1)})=1$. This is achieved when the $\{ | \phi^{(\pm)}_1\rangle_s \}$ states became orthogonal or, equivalently, when the following condition holds:
\begin{equation}
\cos^2(\alpha/2) \cos^2(2\gamma_1)= \sin^2(\alpha/2) \cos^2(2\gamma_2).
\end{equation}
This condition can always be satisfied since the rotation angles $\gamma_1$ and $\gamma_2$ are independent. A maximally entangled state also occurs when the condition $\cos^2(\alpha/2) \sin^2(2\gamma_1)= \sin^2(\alpha/2) \sin^2(2\gamma_2)$. In this last case $E(\Phi^{(2)})=1$, i.e.,  $\{ | \phi^{(\pm)}_2\rangle_s \}$ states became orthogonal.

For the experimental generation of partially entangled states, we replace the type-II BBO crystal by two $0.5$-mm-thick $\mathrm{BBO}$ crystals, cut for type-I phase matching and with their optical axes oriented perpendicular to each other. A half- and quarter-wave plates [HWP and QWP in Fig.~\ref{fig:setup}(a)] in the pump beam allow the preparation of an arbitrary entangled state given by Eq.~(\ref{eq:partial1}) by controlling the parameters $\alpha$ and $\delta$, respectively \cite{Kwiat1}. For this experiment, the HWP$_0$ in the signal arm [see Fig.~\ref{fig:setup}(a)] has been removed and the interferometer was dynamically stabilized as illustrated in Fig.~\ref{fig:setup}(b). This stabilization was done by replacing one of the mirrors at the interferometer by a mirror mounted on a piezoelectric actuator (PZT). Then, a weak laser beam was injected in the interferometer for controlling the MZ fluctuations by means of an electronic circuit. The laser beam wavelength was 632~nm to avoid additional noise. In the experiment, the stability of the MZ interferometer was better than $\lambda/36$.

\begin{figure}[t]
\centering{\rotatebox{-90}{\includegraphics[width=0.35\textwidth]{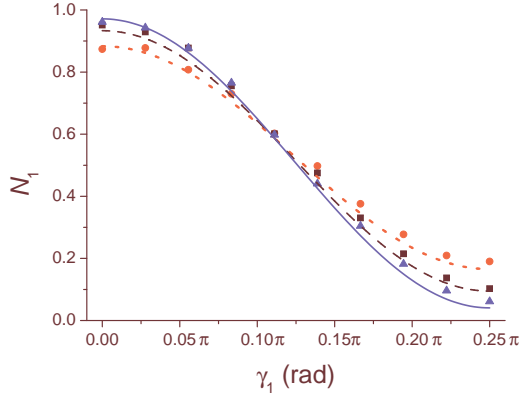}}}
\caption{Measured success probability, $N_1$, of generating output state $|\Psi^{(1)}\rangle$ as a function of the angle $\gamma_1$ at HWP$_1$, while $\gamma_2=\pi/9$, for different values of entanglement of the initial states. Points are experimental results and lines are theoretical prediction using Eq.~(\ref{N1}). $E(\Psi)=0.51$ (triangles and solid line); $E(\Psi)=0.74$ (squares and dashed line); $E(\Psi)=0.90$ (circles and dotted line).}
\label{fig:probabilities2}
\end{figure}

In the first set of measurements we prepared three partially entangled input states by rotating the HWP in the pump beam. For each one we measured the success probability, $N_1$, of generating $|\Phi^{(1)}\rangle$ as a function of $\gamma_1$ while $\gamma_2=\pi/9$. The degree of entanglement of each input state has been obtained through quantum state tomography as described in Ref.~\cite{James}. The values are $E(\Psi) = 0.51\pm 0.02$, $E(\Psi)= 0.74\pm 0.02$ and $E(\Psi)= 0.90\pm 0.02$. To determine $N_1$ experimentally, we measure the coincidence rate between $D_i$ and $D_{s1}$ for the configuration $\gamma_1=\gamma_2=0$ of the interferometer and use this value to normalize the coincidence counts obtained by rotating $\gamma_1$ while $\gamma_2=\pi/9$. The results are shown in Fig.~\ref{fig:probabilities2} and are in good agreement with the theoretical expression given by Eq.~(\ref{N1}). It is seen that for $\gamma_1=\gamma_2=\pi/9$ the three curves intersect each other since in this case the success probability is independent of the input entangled state.

\begin{figure}[t]
\centering{\rotatebox{-90}{\includegraphics[width=0.35\textwidth]{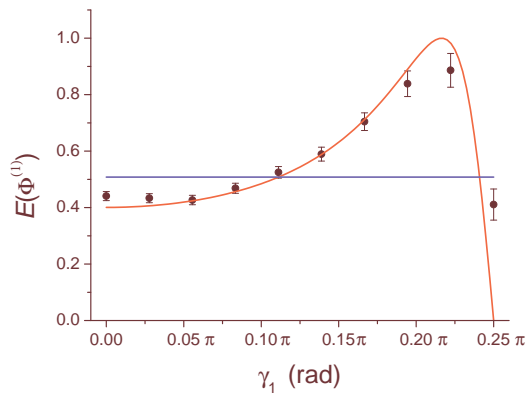}}}
\caption{Measured degree of entanglement of the output state $|\Phi^{(1)}\rangle$ (solid points) as a function of the angle $\gamma_1$ at HWP$_1$, while $\gamma_2=\pi/9$. Here, solid line is the theoretical prediction from Eq.~(\ref{eq:E1}). The straight solid line represents the entanglement of the initial state, $E(\Psi)=0.51$. The probability of generation for each state is given by the triangles in Fig.~\ref{fig:probabilities2}.} \label{fig:entangled}
\end{figure}

For the input state with $E(\Psi) = 0.51$, we also measured the degree of entanglement of the output states $|\Phi^{(1)}\rangle$ as a function of $\gamma_1$ and $\gamma_2=\pi/9$, by means of quantum tomography. Measured degree of entanglement of states $|\Phi^{(1)}\rangle $ appears as points in Fig.~\ref{fig:entangled}. Here, the theoretical prediction for this quantity appears as a solid line using Eq.~(\ref{eq:E1}). As it can be seen there exist a very good agreement. It is also seen there that when $\gamma_1=\gamma_2=\pi/9$, $E(\Psi)=E(\Phi^{(1)})$ and for $\pi/9<\gamma_1<0.24\pi$ the entanglement is concentrated, i.e., $E(\Phi^{(1)})>E(\Psi)$. In the other cases the entanglement decreases. The corresponding probability of generation for each state is given by the triangles in Fig.~\ref{fig:probabilities2}.

Figure~\ref{fig:entangled}, shows that the degree of entanglement of the initial state can be reduced or increased arbitrarily, although probabilistically.  While the experimental concentration of entangled states has already been demonstrated in Refs.~\cite{Gisin,Yamamoto} and used for implementing quantum communication protocols like entanglement swapping \cite{Pan}, in most of the schemes so far, the technique is based on the use of Brewster windows to transmit the polarization entangled twin photons. The entanglement concentration is accomplished because the light which is transmitted through these windows is preferred polarized in one direction, which means that one polarization will have a better transmission while the others are mostly reflected. One can, therefore, concentrate a state like that of Eq.~(\ref{eq:partial}) when $\alpha > \pi/4$ by doing the transmitted light being preferred vertically polarized. However, there is a drawback in this technique because the changes in the degree of entanglement of the transmitted photons are done by adding extra windows to the scheme to change the ratio between the transmitted and reflected lights. Even though this ratio is adjustable it cannot be changed arbitrarily since the transmissions and the reflections at the Brewster windows have distinct dependencies on the material properties used in the window. However, in our scheme a maximally entangled state can always be generated. The only additional issue is that the interferometer being used requires stabilization, but this problem is a minor technical problem which can be easily overcome with a reference laser for doing active stabilization as was done in our experiment. In this case the stabilization circuit allows for having a fixed length difference.

\subsection{Cryptography}

This setup is well suited for an experimental implementation of the so called ``4+2''
quantum key distribution  protocol using single-photon states \cite{Huttner95}. In this
protocol two families of non-orthogonal quantum states are used for establishing a common
secret key between two legitimate users of a quantum channel, Alice and Bob.

The protocol could work as follows: As is described in Sec. \ref{sec:experiment},
the Ar-ion laser pumps a $\mathrm{BBO}$ cut for type-II phase matching.
Photon pairs in a separable polarization state given by $|H\rangle_{s}|V\rangle _{i}$ are selected.
The sender, Alice, rotates the HWP$_0$ at $\gamma_0= \pm \pi/8$ for generating initial states
$|\psi_{\pm} \rangle=(|H\rangle \pm |V\rangle)/\sqrt{2}$. Thus, the output states, after the interferometer,
are given by $| \phi^{(\pm)}_{j} \rangle =\cos (\theta _{j}/2)|H_j\rangle \pm \sin (\theta _{j}/2)|V_j\rangle $,
with $j=1,2$ for the outputs $\left\vert 1 \right\rangle $ and $\left\vert 2\right\rangle $ of the interferometer.
The angles of output states as function of rotation angles of HWP$_1$ and HWP$_2$ are given by:

\begin{eqnarray}
\theta _{1} &=&2\arccos \frac{\cos \left( 2\gamma _{1}\right) }{\sqrt{\cos
^{2}\left( 2\gamma _{1}\right) +\cos ^{2}\left( 2\gamma _{2}\right) }} \\
\theta _{2} &=&2\arccos \frac{\sin \left( 2\gamma _{2}\right) }{\sqrt{\sin
^{2}\left( 2\gamma _{1}\right) +\sin ^{2}\left( 2\gamma _{2}\right) }}
\end{eqnarray}

These states are generated with probabilities $P_{\phi^{(\pm)}_{1}}=
\frac{ \cos(\theta_{2})}{2(\cos((\theta _{1}+\theta _{2})/2)+\cos((\theta _{1}-\theta _{2})/2))}$ and
$P_{\phi^{(\pm)}_{2}}=1-P_{\phi^{(\pm)}_{1}}$ which, under a suitable choice of angles $\theta
_{j}$, can be both equal to $1/2$. Then, Alice randomly chooses which output is sent to
Bob.

The receiver needs to unambiguously discriminate states from pairs
$\{\vert \phi^{(+)}_{1} \rangle , \vert \phi^{(-)}_{1} \rangle \}$ and $\{\vert
\phi^{(+)}_{2} \rangle ,\vert \phi^{(-)}_{2} \rangle \}$. For this purpose Bob must to use
the same interferometric setup depicted in Fig. \ref{fig:setup} as a discrimination device.
For this purpose Bob randomly chooses to rotate HWP$_{1}$ at angles $\gamma _{2}^{(j)}=$ $\frac{1}{2}\arccos
\left( \tan \frac{\theta _{j}}{2}\right)$, which is obtained by considering $\alpha=\theta_j$ and $\beta=\pi/2$
in Eq.(\ref{gamma1}); this allows him to get orthogonal states
$\left\vert \phi^{(\pm) }\right\rangle = \left( \left\vert H_1\right\rangle \pm
\left\vert V_1\right\rangle \right)/\sqrt{2} $ at the output $\left\vert 1\right\rangle $
of Bob's discrimination device, which again is given by Fig. \ref{fig:setup}. Alice
codifies $0$ ($1$) at states $\left\vert \psi^{(+)}\right\rangle$
($\left\vert \psi^{(-)}\right\rangle $), so that Bob obtains an element of the key when the discrimination
angle $\gamma _{2}^{(j)} $ coincides with the output sent by Alice. The output
$\left\vert 2 \right\rangle $ of Bob could be used for detecting the presence of an
eavesdropper, because any intervention of the channel will modify the discrimination
probability.

\section{Summary}
\label{sec:summary}

We have experimentally demonstrated the controlled conclusive modification of non-orthogonal
quantum states. This effective non-unitary evolution of quantum states is generated by coupling
the quantum system to an ancillary system followed by a projective measurement on the ancilla.
We have experimentally proven that the success probability is given by Eqs.~(\ref{eq:psgt}) and
(\ref{eq:psst}). We want to remark that this demonstrate that our protocol is optimal for
the class of states considered. This setup is well suited for an experimental implementation of
the so called ``4+2'' \cite{Huttner95} quantum key distribution protocols using one-photon states,
where two families of non-orthogonal quantum states are used for establishing a common
secret key between two legitimate users of a quantum channel, Alice and Bob.

This experimental setup is also well suited for the controlled generation of polarization mixed states
of one and two photons by changing in a controlled way the length of one of the arms of the interferometer
within the coherence length of downconverted photons. This would also allow to experimentally study the
discrimination of two mixed states.

The generalization of the results here reported and the case of higher dimensional systems seems feasible
considering recent advances in integrated waveguide quantum circuits  \cite{Politi2008} and considering
protocols like the one proposed by Jim\'enez \textit{et al.} \cite{Jimenez}.

\section{Acknowledgment}

This work was supported by Grants Milenio ICM P06-067F and
FONDECyT 1061046 and 1080383. F.A. Torres-Ruiz thanks to MECESUP for scholarship support.

\end{document}